\documentclass[12pt,draftcls,onecolumn]{IEEEtran}

\def \be {\begin{equation}}
\def \ee {\end{equation}}

\def \nn {\nonumber}

\usepackage{ifpdf}

\usepackage[dvips]{graphicx}
\usepackage{color}

\usepackage[cmex10]{amsmath}
\usepackage{amssymb}

\begin{document}
%
\title{\LARGE \bf Passivity-Based Adaptive Control for Visually Servoed Robotic Systems}
%
%
%

\author{Hanlei~Wang 
\thanks{The author is with the Science and Technology on Space Intelligent Control
Laboratory, Beijing Institute of Control Engineering,
Beijing 100190, China (e-mail: hlwang.bice@gmail.com).}
}
\maketitle

\begin{abstract}
This paper investigates the visual servoing problem for robotic systems with uncertain kinematic, dynamic, and camera parameters. We first present the passivity properties associated with the overall kinematics of the system, and then propose two passivity-based adaptive control schemes to resolve the visual tracking problem. One scheme employs the adaptive inverse-Jacobian-like feedback, and the other employs the adaptive transpose Jacobian feedback. With the Lyapunov analysis approach, it is shown that under either of the proposed control schemes, the image-space tracking errors converge to zero without relying on the assumption of the invertibility of the estimated depth. Numerical simulations are performed to show the tracking performance of the proposed adaptive controllers.
\end{abstract}


\section{Introduction}

The interests in visual servoing for robots have lasted for many years (see, e.g., \cite{Espiau1992_TRA,Hutchinson1996_TRA,Malis2002_TRA,Astolfi2002_TRA,Hamel2002_TRA,Mahony2002_ICRA,Liu2006_TRO,Fujita2007_TCST,Hu2009_TAC,Cheah2010_Aut}). The visual servoing schemes can roughly be classified into two categories (see, e.g., \cite{Hutchinson1996_TRA}): position-based scheme and image-based scheme. The familiar advantage of the
image-based servoing scheme may be that the possible errors in camera modeling
and calibration are avoided, and that the reduction of the error in the image space implies that of the
error in the physical task space (or Cartesian space). However, the direct use of image features in feedback control complicates the kinematics of the robotic system, and furthermore parametric uncertainty often arises (see, e.g., \cite{Liu2006_TRO}).

For handling the nonlinearity and parametric uncertainty of the models of the visually servoed robotic systems, many model-based adaptive control schemes are proposed, e.g., \cite{Cheah2006_IJRR,Liu2006_Aut,Liu2006_TRO,Dixon2007_TAC,Wang2007_TRO,Cheah2007_ICRA,Cheah2010_Aut,Leite2011_IFAC,Wang2012b_Mech,Lizarralde2013_Aut,Wang2015_AUT,Wang2014AdaptiveJacobian_arXiv}.
The work in \cite{Akella2005_TRO,Cheah2006_IJRR,Liu2006_Aut,Lizarralde2013_Aut,Wang2014AdaptiveJacobian_arXiv} studies the visual tracking problem under the assumption that the depth is constant, in which case, the overall Jacobian matrix that describes the relation between the joint-space velocity and the image-space velocity is linearly parameterized (see, e.g., \cite{Cheah2006_IJRR,Astolfi2002_TRA}). The results that explicitly take into consideration the time-varying depth information of the camera appear in \cite{Liu2006_TRO,Wang2007_TRO,Cheah2007_ICRA,Cheah2010_Aut,Leite2011_IFAC,Wang2012b_Mech,Wang2015_AUT}, and as is demonstrated in, e.g., \cite{Liu2006_TRO,Wang2007_TRO,Cheah2010_Aut}, the overall Jacobian matrix in this case cannot be expressed as the linearity-in-parameters form since the uncertain depth acts as the denominator in the overall Jacobian matrix. The adaptive schemes in \cite{Liu2006_TRO,Cheah2010_Aut}, by adaptation to the system uncertainty, ensure that the image-space position is regulated to the desired one asymptotically. The adaptive schemes in \cite{Wang2007_TRO,Cheah2007_ICRA,Leite2011_IFAC,Wang2012b_Mech,Wang2015_AUT} realize image-space trajectory tracking regardless of the system uncertainty (it is noted that the work in \cite{Leite2011_IFAC} confines to the case of the target object with the specific spherical geometry so as to exploit certain invariant quantities, limiting its applications). The tracking control schemes in the existing work (e.g., \cite{Wang2007_TRO,Cheah2007_ICRA,Wang2012b_Mech,Wang2015_AUT}), however, rely on the assumption of invertibility of the estimated depth (or the use of parameter projection to guarantee this) to ensure the tracking error convergence, due in part to the inadequate exploitation of the (potentially beneficial) structural property of the overall kinematics.

In this paper, we start from formulating a {new form of passivity associated with the overall kinematics} of the visually servoed robotic system, based on which, we present two adaptive controllers {for 3-dimensional visual tracking} that neither relies on the assumption of the invertibility of the estimated depth nor the use of parameter projection algorithm to ensure its invertibility, in contrast to \cite{Wang2007_TRO,Cheah2007_ICRA,Wang2012b_Mech}. {The avoidance of this assumption or parameter projection is important in that no a priori knowledge of the depth information (used for calculating the parameter region) is required and in addition, we do not need to concern where the estimated depth parameter finally stays}. Among the two controllers, one employs the adaptive inverse-Jacobian-like feedback and the other employs the adaptive transpose Jacobian feedback. Our work extends the case of the constant depth considered in \cite{Wang2014AdaptiveJacobian_arXiv} (adaptive inverse Jacobian control) and \cite{Cheah2006_IJRR} (adaptive transpose Jacobian control) to that of the time-varying depth, by exploiting the depth-related passivity of the overall kinematics and incorporating adaptation to the uncertain depth. We also show that one reduced version of the adaptive inverse-Jacobian-like controller (referred to here as a separation approach due to its separation property, which is in contrast to the dynamic scheme in \cite{Leite2011_IFAC} that relies on a target object with the specific spherical geometry, persistent excitation condition, and an additional Cartesian-space sensor) is a qualified adaptive kinematic scheme that fits well for robots having an unmodifiable joint servoing controller yet admitting the the design of the joint velocity command (e.g., most industrial robots)---see Remark 2, which is in contrast to the existing kinematic schemes (e.g., \cite{Espiau1992_TRA,Hutchinson1996_TRA}) that lack adequation consideration of the robot dynamics. While the adaptive transpose Jacobian controller can be considered as a special case of \cite{Wang2015_AUT}, the depth-related passivity and the adaptive inverse-Jacobian-like controller with the separation property constitute the contribution of our result with respect to \cite{Wang2015_AUT}.

In summary, the major contribution of this work is that {the passivity properties associated with the overall kinematics are explicitly presented, and that two adaptive controllers are proposed and shown to be convergent without the need of the assumption that the estimated depth is invertible; in addition, the separation property of the adaptive inverse-Jacobian-like controller yields an adaptive kinematic controller applicable to most industrial robots. It may be worth remarking that for {most image-space tracking tasks (i.e., the desired image-space velocity is not identically zero at the final state)}, the invertibility of the estimated depth (at the final state) is required and can be ensured by the proposed controllers, but most existing results cannot ensure this and the common practice is to rely on assumption or use a relatively complex projection algorithm (requiring certain a priori information of the depth and the determination of an appropriate parameter region)}. A preliminary version of the paper was presented in \cite{Wang2014_AUCC} where the passivity of the overall kinematics and adaptive transpose Jacobian control were presented, and here we expand this version to additionally cover the adaptive inverse-Jacobian-like control.

\section{Kinematics and Dynamics}

In this paper, we consider a visually servoed robotic system that consists of an $n$-DOF (degree-of-freedom) manipulator and a standard fixed pinhole camera (see, e.g., \cite{Forsyth2012_Book}), where the manipulator end-effector motion is mapped to the image space by the camera. For the convenience of the theoretical formulation, the number of the feature points that are under consideration is determined as one\footnote{{The consideration of one feature point here is for the sake of convenience of theoretical formulation, and the extension to the case of multiple feature points with different depths can be directly performed in a way similar to \cite{Wang2015_AUT}. For instance, in the case of three feature points, we can stack the image-space positions of the three feature points $x_1\in R^2$, $x_2\in R^2$, and $x_3\in R^2$ as a single vector $x=\left[x_1^T,x_2^T,x_3^T\right]^T$ and the image Jacobian matrices associated with these feature points as a single matrix. The ensuing procedure would then be straightforward. It may be worth noting that in this case, the depths are indirectly controlled by controlling sufficient number of feature points.}}. 

\subsection{Kinematics}

Let $x\in R^2$ and $r\in R^3$, respectively, denote the position of the projection of the feature point on the image plane and the position of the feature point with respect to the base frame of the manipulator. The mapping from $r$ to $x$ can be written as \cite{Forsyth2012_Book,Liu2006_TRO}
\be
\begin{bmatrix}
x\\
1
\end{bmatrix}=\frac{1}{z(q)}{\mathcal H}\begin{bmatrix}
r\\
1
\end{bmatrix}
\ee
where ${\mathcal H}=\left[D, b\right]\in R^{3\times 4}$ (with $D\in R^{3\times3}$ and $b\in R^3$) is the perspective projection matrix, $q\in R^n$ denotes the joint position of the manipulator, and $z(q)=d_3^T r+b_3\in R$ with $b_3$ being the third element of $b$ and $d_3^T$ being the third row of $D$ denotes the depth of the feature point with respect to the camera frame.  The relationship between the image-space velocity $\dot x$ and the feature-point velocity $\dot r$ can be written as \cite{Liu2006_TRO}
\be
\label{eq:1} \dot {x} =\frac{1}{z(q)}\left(\bar D-x d_3^T\right) \dot r
\ee
where $\bar D\in R^{2\times 3}$ is composed of the first two rows of $D$, and $N(x)=\bar D-x d_3^T\in R^{2\times 3}$ is the depth-independent interaction matrix defined by \cite{Liu2006_TRO}. Obviously, the time derivative of the depth can be written as $\dot z(q)=d_3^T \dot r$ \cite{Liu2006_TRO}. Furthermore, it is assumed that {the depth $z(q)$ is uniformly positive during the motion of the manipulator}.

Let $r_0\in R^3$ denote the position of a reference point on the end-effector with respect to the manipulator base frame, $\dot r_0$  its translational velocity, and $\omega_0\in R^3$ the angular velocity of the end-effector expressed in the manipulator base frame. The velocities $\dot r_0$ and  $\omega_0$ relate to the joint velocity $\dot q$ by \cite{Craig2005_Book,Spong2006_Book}
\be
\begin{bmatrix}
\dot r_0\\
\omega_0
\end{bmatrix}=J_r(q)\dot q
\ee
where $J_r(q)\in R^{6\times n}$ denotes the manipulator Jacobian matrix.

The relationship between the feature-point velocity $\dot{r}$ and the joint velocity $\dot q$ can be written as \cite{Wang2007_TRO} (see also \cite{Hutchinson1996_TRA,Craig2005_Book,Spong2006_Book})
\be
\label{eq:3}
\dot{r}=\underbrace{\begin{bmatrix}I_3 & -S(c)\end{bmatrix}}_{J_f}J_r(q)\dot{q}
\ee
where $I_3$ is the $3\times 3$ identity matrix, $c=\left[c_1,c_2,c_3\right]^T\in R^3$ denotes the position of the feature point with respect to the reference point on the manipulator end-effector expressed in the manipulator base frame, and the skew-symmetric form $S(c)$ is defined as $$S(c)=\begin{bmatrix}0 & -c_3 & c_2\\
c_3 &0& -c_1\\
-c_2& c_1 & 0\end{bmatrix}.$$

Combining (\ref{eq:1}) and (\ref{eq:3}) yields the following overall kinematic equation \cite{Liu2006_TRO,Cheah2010_Aut}
\be
\label{eq:4}
\dot{x}=\frac{1}{z(q)}\underbrace{N(x)J_f J_r(q)}_{J(q,x)}\dot{q}
\ee
where $J(q,x)$ is a Jacobian matrix. The structural property of (\ref{eq:1}) allows us to decompose $J(q,x)$ as
\be
\label{eq:5}
J(q,x)=\underbrace{\bar D J_f J_r(q)}_{J_z^\perp(q)}-x \underbrace{d_3^TJ_f J_r(q)}_{J_z(q)}
\ee
where $J_z^\perp(q)$ is a Jacobian matrix that maps the joint velocity $\dot q$ to a plane which is parallel to the image plane (i.e., perpendicular to the depth direction), and $J_z(q)$ is a Jacobian matrix that describes the relation between the changing rate of the depth $z(q)$ and $\dot q$ (see, e.g., \cite{Liu2006_TRO}), i.e.,
\be
\label{eq:6}
\dot z(q)=J_z(q)\dot q.
\ee
We note that whether the depth $z(q)$ is time-varying or not, the Jacobian matrix $J_z^\perp(q)$ is contained in $J(q,x)$. For this, as in \cite{Wang2015_AUT}, we refer to $J_z^\perp(q)$ as the \emph{depth-rate-independent Jacobian matrix}.

The overall kinematics (\ref{eq:4}) has the following property.

\emph{Property 1:} For an arbitrary vector $\phi\in R^2$, the two quantities $z(q)\phi$ and $\dot z(q)\phi$ depend linearly on a constant depth parameter vector $a_z\in R^{m_1}$ \cite{Liu2006_TRO,Cheah2010_Aut}, i.e.,
\begin{align}
z(q)\phi=&Y_z(q,\phi)a_z\\
\dot z(q)\phi=& \bar Y_z(q,\dot q,\phi)a_z
\end{align}
which also directly yields
\be
\label{eq:9}
\phi J_z(q)\dot q=\dot z(q)\phi=\bar Y_z(q,\dot q,\phi)a_z
\ee
where $Y_z(q,\phi)\in R^{2\times m_1}$ and $\bar Y_z(q,\dot q,\phi)\in R^{2\times m_1}$ are regressor matrices. In addition, $J(q,x)\dot q$ can be linearly parameterized \cite{Cheah2010_Aut}, which thus leads to
\be
\label{eq:10}
J_z^\perp(q)\xi=Y_z^\perp(q,\xi)a_z^\perp
\ee
where $a_z^\perp\in R^{m_2}$ is the unknown depth-rate-independent parameter vector, $\xi\in R^n$ is a vector, and $Y_z^\perp(q,\xi)\in R^{2\times m_2}$ is the depth-rate-independent kinematic regressor matrix. 

\subsection{Dynamics}

The dynamics of the $n$-DOF manipulator can be written as \cite{Slotine1991_Book,Spong2006_Book}
\be
\label{eq:12}
M(q)\ddot q+C(q,\dot q)\dot q+g(q)=\tau
\ee
where $M(q)\in R^{n\times n}$ is the inertia matrix, $C(q,\dot q)\in R^{n\times n}$ is the Coriolis and centrifugal matrix, $g(q)\in R^n$ is the gravitational torque, and $\tau\in R^n$ is the joint control torque. In this paper, we assume that the number of the DOFs of the manipulator is not less than two, i.e., $n\ge 2$. Three familiar properties associated with the dynamic model (\ref{eq:12}) that shall be useful for the subsequent controller design and stability analysis are listed as follows (see, e.g., \cite{Slotine1991_Book,Spong2006_Book}).

\emph{Property 2:} The inertia matrix $M(q)$ is symmetric and uniformly positive definite.

\emph{Property 3:} The Coriolis and centrifugal matrix $C(q,\dot q)$ can be appropriately chosen such that $\dot M(q)-2C(q,\dot q)$ is skew-symmetric.

\emph{Property 4:} The dynamic model (\ref{eq:12}) depends linearly on a constant dynamic parameter vector $a_d\in R^{p}$, thus yielding
\be
M(q)\dot \zeta+C(q,\dot q)\zeta+g(q)=Y_d(q,\dot{q},\zeta,\dot \zeta)a_d
\ee
where $Y_d(q,\dot{q},\zeta,\dot \zeta)\in R^{n\times p}$ is the regressor matrix, $\zeta\in R^n$ is a differentiable vector, and $\dot\zeta$ is the time derivative of $\zeta$.

\section{Adaptive Control}

In this section, we aim to design adaptive controllers for the visually servoed robotic system by formulating and exploiting the passivity of the overall kinematics. The control objective is to ensure the converge of the image-space tracking errors, i.e., $x-x_d\to 0$ and $\dot x-\dot x_d\to 0$ as $t\to\infty$, where $x_d\in R^2$ denotes the desired image-space position and it is assumed that $x_d$, $\dot x_d$, and $\ddot x_d$ are all bounded.

\subsection{Passivity Associated With the Overall Kinematics}

Combining (\ref{eq:4}) and (\ref{eq:5}), we can rewrite the overall kinematics (\ref{eq:4}) as
\be
\label{eq:14}
z(q)\dot x+\frac{1}{2}\dot z(q)x=\underbrace{\left[J_z^\perp(q) -\frac{1}{2}x J_z(q)\right]\dot q}_{u}
\ee
where $u$ is a virtual or intermediate control input.

\emph{Proposition 1:} The system (\ref{eq:14}) is passive with respect to the input-output pair $(u,x)$.

\emph{Proof:} Consider the following function (which is actually one part of the Lyapunov function in \cite{Liu2006_TRO})
\be
V_s=\frac{z(q)}{2}x^T x.
\ee
Differentiating $V_s$ with respect to time along the trajectories of (\ref{eq:14}) yields
\be
\dot V_s=x^T u
\ee
which can be rewritten as
\be
\int_0^t x^T(r)u(r)dr=V_s(t)-V_s(0)\ge -V_s(0).
\ee
This implies that the system (\ref{eq:14}) is passive with respect to the input-output pair $(u,x)$ in the sense of \cite{Lozano2000_Book}. \hfill {\small $\blacksquare$}

According to the standard passivity-based design methodology \cite{Lozano2000_Book}, a simple output feedback for $u$ can result in the convergence of the output $x$ to the origin (the case of nonzero equilibrium shall be similar). {The regulation algorithm of \cite{Liu2006_TRO} can be considered as a combined application of the passivity of the overall kinematics here and the standard passivity of the manipulator dynamics (see, e.g., \cite{Spong2006_Book}). The passivity concerning the overall kinematics has also been examined in \cite{Hamel2002_TRA}, yet the storage function in \cite{Hamel2002_TRA} is independent of the depth while the storage function considered here is explicitly related to the depth. {The main benefit of introducing this depth-related passivity}, as is shown later, is the avoidance of the restrictive assumption of the invertibility of the estimated depth without relying on parameter projection.}

The control above, however, is not enough for realizing the objective of image-space tracking, in which case, it is expected to drive the tracking error $\Delta x=x-x_d$ to the origin. To this end, we would like to apply the feedback passivation strategy \cite{Byrnes1991_TAC}, i.e., let the control $u$ be given by
\be
\label{eq:18}
u=\bar u+\underbrace{z(q)\dot x_d+\frac{1}{2}\dot z(q)x_d}_\text{feedback passivation}
\ee
where $\bar u$ becomes the new virtual control input.
Substituting (\ref{eq:18}) into (\ref{eq:14}) gives
\be
\label{eq:19}
z(q)\Delta \dot x+\frac{1}{2}\dot z(q)\Delta x=\bar u.
\ee

\emph{Proposition 2:} The system (\ref{eq:19}) is passive with respect to the input-output pair $(\bar u,\Delta x)$.

The proof of Proposition 2 shall be similar to that of Proposition 1. 

Using (\ref{eq:6}), we can rewrite (\ref{eq:18}) as
\be
\label{eq:20}
\bar u=-z(q)\dot x_d+\Big[\underbrace{J_z^\perp(q)-\frac{x+x_d}{2}J_z(q)}_{J^\ast}\Big]\dot q.
\ee
Obviously, if the Jacobian matrix $J^\ast$ has full row rank, the virtual control $\bar u$ can be realized by the joint velocity $\dot q$. 


\subsection{Adaptive Inverse-Jacobian-like Control}

Let us now start the adaptive controller design based on the passivity enjoyed by the overall kinematic equation.

Due to the passivity of the input-output pair $(\bar u, \Delta x)$, the standard passivity-based design \cite{Lozano2000_Book} suggests that the virtual control $\bar u=-\bar K \Delta x$ with $\bar K$ being a symmetric positive definite matrix would be qualified for realizing the image-space tracking, yet, not necessarily give guaranteed performance due to the variation of the depth $z(q)$. To accommodate the varying and uncertain depth, we propose the following virtual control
\be
\label{eq:21}
\bar u=-\alpha \hat z(q)\Delta x
\ee
where $\alpha>0$ is a design constant and $\hat z(q)$ is the estimate of $z(q)$ which is obtained by replacing $a_z$ in $z(q)$ with its estimate $\hat a_z$. The use of the virtual control (\ref{eq:21}) is inspired by the performance guaranteed adaptive control for robot manipulators in \cite[Sec.~3.2]{Slotine1989_Aut}. However, it should be emphasized that the virtual control (\ref{eq:21}) is not the actual control since it does not take into account the dynamic effect of the manipulator.

Keeping (\ref{eq:21}) in mind and based on (\ref{eq:20}), we define a joint reference velocity using the estimated Jacobian matrix as
\be
\label{eq:22}
\dot q_r=\hat J^{\ast +}\hat z(q)\dot x_r
\ee
where $\hat J^\ast$ is the estimate of $J^\ast$ which is obtained by replacing $a_z^\perp$ and $a_z$ in $J^\ast$ with their estimates $\hat a_z^\perp$ and $\hat a_z$, respectively, $\hat J^{\ast +}=\hat J^{\ast T} \big(\hat J^\ast \hat J^{\ast T}\big)^{-1}$ is the standard generalized inverse of $\hat J^\ast$ (see, e.g., \cite{Cheah2006_TAC}), and $\dot x_r=\dot x_d-\alpha\Delta x$. Differentiating (\ref{eq:22}) with respect to time yields the joint reference acceleration
\begin{align}
\label{eq:23}
\ddot q_r=&\hat J^{\ast+}\left[\hat z(q)\ddot x_r+\dot{\hat z}(q) \dot x_r-\dot{\hat J}^\ast \dot q_r\right]\nn\\
&+(I_n-\hat J^{\ast+}\hat J^\ast)\dot{\hat J}^{\ast T} \hat J^{\ast+ T}\dot q_r
\end{align}
where the standard result concerning the time derivative of $\hat J^{\ast +}$ is used and $I_n$ is the $n\times n$ identity matrix.

Let us now define a sliding vector as
\be
\label{eq:24}
s=\dot q-\dot q_r
\ee
whose derivative with respect to time can be written as
\be
\dot s=\ddot q -\ddot q_r.
\ee

Premultiplying both sides of (\ref{eq:24}) by $ J^\ast$ and using (\ref{eq:4}), (\ref{eq:6}), (\ref{eq:22}), and Property 1 yields
\begin{align}
\label{eq:26}
J^\ast s=&z(q)\dot x+\frac{1}{2}\dot z(q)\Delta x-\hat J^\ast \dot q_r\nn\\
&+Y_z^\perp(q,\dot q_r)\Delta a_z^\perp-\frac{1}{2}\bar Y_z(q,\dot q_r,x+x_d)\Delta a_z\nn\\
=& z(q)\left(\Delta \dot x+\alpha \Delta x\right)+\frac{1}{2}\dot z(q)\Delta x+Y_z^\perp(q,\dot q_r)\Delta a_z^\perp\nn\\
&-\big[\underbrace{Y_z(q,\dot x_r)+\frac{1}{2}\bar Y_z(q,\dot q_r,x+x_d)}_{Y_z^\ast(q,\dot q_r,x+x_d,\dot x_r)}\big]\Delta a_z
\end{align}
where $\Delta a_z^\perp=\hat a_z^\perp-a_z^\perp$ and $\Delta a_z=\hat a_z-a_z$. Equation (\ref{eq:26}) can be rewritten as
\begin{align}
\label{eq:27}
z(q)\Delta \dot x+\frac{1}{2}\dot z(q)\Delta x=&-\alpha z(q)\Delta x-Y_z^\perp(q,\dot q_r)\Delta a_z^\perp\nn\\
&+Y_z^\ast(q,\dot q_r,x+x_d,\dot x_r)\Delta a_z+J^\ast s.
\end{align}

Now we propose the following control law
\be
\label{eq:28}
\tau=-K s+Y_d(q,\dot q,\dot q_r,\ddot q_r)\hat a_d
\ee
where $\hat a_d$ is the estimate of $a_d$ and $K$ is a symmetric positive definite matrix.
The adaptation laws for updating $\hat a_d$, $\hat a_z$, and $\hat a_z^\perp$ are given as
\begin{align}
\label{eq:29}
\dot{\hat a}_d=&-\Gamma_d Y_d^T(q,\dot q,\dot q_r,\ddot q_r) s\\
\label{eq:30}
\dot{\hat a}_z=&-\Gamma_zY_z^{\ast T}(q,\dot q_r,x+x_d,\dot x_r)\Delta x\\
\label{eq:31}
\dot{\hat a}_z^\perp=&\Gamma_z^\perp Y_z^{\perp T}(q,\dot q_r)\Delta x
\end{align}
where $\Gamma_d$, $\Gamma_z$, and $\Gamma_z^\perp$ are symmetric positive definite matrices.

\emph{Remark 1:} The feedback term $-Ks$ in (\ref{eq:28}) can be interpreted as inverse-Jacobian-like control based on (\ref{eq:27}), and it appears that both the image-space tracking errors and parameter estimation errors $\Delta a_z$ and $\Delta a_z^\perp$ are included. The parameter adaptation laws (\ref{eq:30}) and (\ref{eq:31}) rely on the two regressor matrices that use the joint reference velocity $\dot q_r$ and thus are actually adaptive in the sense that they are updated in accordance with the updating of the parameter estimates. The avoidance of the assumption of invertibility of the estimated depth is reflected in (\ref{eq:27}). Asymptotically, the final closed kinematic loop behaves like the one with a feedforward $z(q)\dot x_d$ and a feedback $-\alpha z(q)\Delta x$ since the term related to the parameter estimation errors $\Phi=-Y_z^\perp(q,\dot q_r)\Delta a_z^\perp+Y_z^\ast(q,\dot q_r,x+x_d,\dot x_r)\Delta a_z$ converges to zero asymptotically (which can be shown by the consequence of Theorem 1 below). From the result that $\Phi\to 0$ as $t\to\infty$, we have that
\begin{align}
&-\left(\hat J_z^\perp-J_z^\perp\right)\hat J^{\ast +}\hat z(q)\dot x_d+[\hat z(q)-z(q)]\dot x_d\nn\\
&+\frac{x+x_d}{2}\left[\hat {\dot z}(q)-\dot z(q)\right]\hat J^{\ast +}\hat z(q)\dot x_d\to 0
\end{align}
as $t\to\infty$. Then, it can be shown by contradiction that $\hat z(q)\dot x_d \ne 0$ so long as $z(q)\dot x_d\ne 0$ or $\dot x_d\ne 0$ as $t\to\infty$. This can be interpreted as ``if $\dot x_d$ does not converge to zero, then the estimated depth $\hat z(q)$ would converge to an invertible quantity''.

Substituting the control law (\ref{eq:28}) into (\ref{eq:12}) gives
\be
\label{eq:32}
M(q)\dot s+C(q,\dot q)s=-K  s+Y_d(q,\dot q,\dot q_r,\ddot q_r)\Delta a_d
\ee
where $\Delta a_d=\hat a_d-a_d$ is the dynamic parameter estimation error.

The closed-loop behavior of the visually servoed robotic system can then be described by
(\ref{eq:27}), (\ref{eq:32}), (\ref{eq:29}), (\ref{eq:30}), and (\ref{eq:31}).

We are presently ready to formulate the following theorem.

\emph{Theorem 1:} For the visually servoed robotic system given by (\ref{eq:4}), (\ref{eq:6}), and (\ref{eq:12}), the control law (\ref{eq:28}) and the parameter adaptation laws (\ref{eq:29}), (\ref{eq:30}), and (\ref{eq:31}) ensure the convergence of the image-space tracking errors, i.e., $\Delta x\to 0$ and $\Delta \dot x\to 0$ as $t\to\infty$.

\emph{Proof:} Following \cite{Slotine1987_IJRR,Ortega1989_Aut}, we take into account the Lyapunov-like function candidate $
V_1=(1/2)s^T M(q)s+(1/2) \Delta a_d^T \Gamma_d^{-1}\Delta a_d
$, and differentiating $V_1$ with respect to time along the trajectories of (\ref{eq:32}) and (\ref{eq:29}) and exploiting Property 3, we have $\dot V_1=-s^TK s\le 0$, which then implies that $s\in  {\cal L}_2\cap {\cal L }_\infty$ and $\hat a_d\in {\cal L}_\infty$.

The boundedness of $J^\ast$ implies that $J^\ast s\in {\cal L}_2$. In addition, $z(q)$ is uniformly positive by assumption. Then, there must exist a positive constant $l_M$ such that $\int_0^t (1/z(q(r))) s^T(r) J^{\ast T}(r) J^\ast(r) s(r)dr\le l_M$, $\forall t\ge 0$. Based on the passivity of the system kinematics, consider a nonnegative function
\begin{align}
\label{eq:33}
V_2=&\frac{1}{2}z(q)\Delta x^T \Delta x+\frac{1}{2}\Delta a_z^T \Gamma_z^{-1}\Delta a_z+\frac{1}{2}\Delta a_z^{\perp T}\Gamma_z^{\perp -1}\Delta a_z^\perp\nn\\
&+\frac{1}{2\alpha}\bigg[\underbrace{l_M-\int_0^t \frac{1}{z(q(r))} s^T(r) J^{\ast T}(r) J^\ast(r) s(r)dr}_{\Pi^\ast}\bigg]
\end{align}
where the term $\Pi^\ast$ follows the result in \cite[p.~118]{Lozano2000_Book}. Differentiating $V_2$ with respect to time along the trajectories of (\ref{eq:27}), (\ref{eq:30}), and (\ref{eq:31}) gives
\be
\label{eq:34}
\dot V_2=-\alpha z(q)\Delta x^T \Delta x+\Delta x^T J^\ast s-\frac{1}{2\alpha z(q)}s^T J^{\ast T}J^\ast s.
\ee
Combining the following result derived from the standard inequality
\be
\Delta x^T J^\ast s\le \frac{\alpha z(q)}{2}\Delta x^T \Delta x+\frac{1}{2\alpha z(q)}s^T  J^{\ast T} J^\ast s
\ee
and (\ref{eq:34}) yields
\be
\dot V_2\le -\frac{\alpha z(q)}{2}\Delta x^T \Delta x\le 0.
\ee
This implies that $\Delta x\in {\cal L}_2\cap {\cal L}_\infty$, $\hat a_z\in {\cal L}_\infty$, and $\hat a_z^\perp\in {\cal L}_\infty$. Then, we get the result that $x\in {\cal L}_\infty$, $\hat z(q)\in {\cal L}_\infty$, and $\dot x_r\in {\cal L}_\infty$. From (\ref{eq:22}), we obtain that $\dot q_r\in {\cal L}_\infty$ if $\hat J^\ast$ has full row rank (which ensures the existence of the generalized inverse of $\hat J^\ast$ according to the standard matrix theory). Therefore, $\dot q=s+\dot q_r\in{\cal L}_\infty$. From the overall kinematics (\ref{eq:4}), we have that $\dot x\in {\cal L}_\infty$ and further $\Delta \dot x\in {\cal L}_\infty$, which then implies that $\Delta x$ is uniformly continuous. From the properties of square-integrable and uniformly continuous functions \cite[p.~232]{Desoer1975_Book}, we have that $\Delta x\to 0$ as $t\to\infty$. From (\ref{eq:30}) and (\ref{eq:31}), we obtain that $\dot{\hat a}_z\in {\cal L}_\infty$ and $\dot{\hat a}_z^\perp\in {\cal L}_\infty$, giving rise to the boundedness of $\dot{\hat z}(q)$ and $\dot {\hat J}^\ast$. Then, we obtain from (\ref{eq:23}) that $\ddot q_r\in {\cal L}_\infty$. Based on (\ref{eq:32}) and the fact that $M(q)$ is uniformly positive definite (by Property 2), we have that $\dot s\in {\cal L}_\infty$. This immediately implies that $\ddot q\in{\cal L}_\infty$. From the differentiation of (\ref{eq:4}) with respect to time, we have that $\ddot x\in {\cal L}_\infty$. Hence, $\Delta \ddot x\in {\cal L}_\infty$, which means that $\Delta \dot x$ is uniformly continuous. From Barbalat's Lemma \cite{Slotine1991_Book}, we obtain that $\Delta \dot x\to 0$ as $t\to\infty$. \hfill {\small $\blacksquare$}

\subsection{Adaptive Transpose Jacobian Control}

The adaptive transpose Jacobian control is given as
\begin{align}
\label{eq:35}
\tau=&-\hat J^{\ast T}K_1 \hat J^\ast s+Y_d(q,\dot q,\dot q_r,\ddot q_r)\hat a_d\\
\label{eq:36}
\dot{\hat a}_d=&-\Gamma_d Y_d^T(q,\dot q,\dot q_r,\ddot q_r) s\\
\label{eq:37}
\dot{\hat a}_z=&-\Gamma_zY_z^{\ast T}(q,\dot q,x+x_d,\dot x_r)\Delta x\\
\label{eq:38}
\dot{\hat a}_z^\perp=&\Gamma_z^\perp Y_z^{\perp T}(q,\dot q)\Delta x
\end{align}
where $K_1$ is a symmetric positive definite matrix. This controller turns out to be actually identical to a reduced version of the one in \cite{Wang2015_AUT} (i.e., by assuming
that the image-space velocity can be precisely obtained). Detailed analysis can be found in our preliminary work \cite{Wang2014_AUCC}. The difference between the adaptive transpose Jacobian control scheme and the adaptive inverse-Jacobian-like control scheme not only lies in the feedback part but in the depth and depth-rate-independent kinematic parameter adaptation laws. In fact, the regressor matrices used in (\ref{eq:37}) and (\ref{eq:38}) are not adaptive in contrast with the adaptive ones used in the adaptive inverse-Jacobian-like control scheme. The expense that we have to pay due to the use of non-adaptive regressor matrices is a relatively strong feedback, i.e., the adaptive transpose Jacobian feedback $-\hat J^{\ast T}K_1 \hat J^\ast s$ in (\ref{eq:35}).

\emph{Remark 2:}
\begin{enumerate}

\item In most industrial robotic applications, the available control command is the joint velocity (position) rather than the joint torque. {It seems interesting that one reduced version of the proposed adaptive inverse-Jacobian-like control does fit this scenario well, i.e., the adaptive kinematic control scheme given by [from (\ref{eq:22}), (\ref{eq:30}), and (\ref{eq:31})]
    \be
    \label{eq:39}
    \begin{cases}
    \dot q_r=\hat J^{\ast +}\hat z(q)\dot x_r\\
   \dot{\hat a}_z=-\Gamma_zY_z^{\ast T}(q,\dot q_r,x+x_d,\dot x_r)\Delta x\\
\dot{\hat a}_z^\perp=\Gamma_z^\perp Y_z^{\perp T}(q,\dot q_r)\Delta x
    \end{cases}
    \ee where $\dot q_r$ acts as the joint velocity command.} This kinematic control scheme yields a closed-loop system given by (\ref{eq:27}) and the two adaptation laws in (\ref{eq:39}). Under the common assumption that the joint servoing module guarantees that the joint velocity tends sufficiently fast to the joint velocity command [i.e., the joint reference velocity $\dot q_r$ given in (\ref{eq:39})] in the sense that $s$ is square-integrable and bounded, the term $J^\ast s$ in (\ref{eq:27}) is square-integrable and bounded. Then, taking the same nonnegative function as (\ref{eq:33}) and following similar analysis as in the proof of Theorem 1 would immediately yield the conclusion that the image-space tracking errors converge to zero. 

\item The adaptive transpose Jacobian control, unfortunately, does not enjoy the above properties, which is mainly due to the transpose Jacobian feedback in (\ref{eq:35}) and the parameter adaptation laws (\ref{eq:37}) and (\ref{eq:38}).

\end{enumerate}

\section{Simulation Results}

Consider a visually servoed robotic system composed of a standard three-DOF manipulator and a fixed camera (as shown in Fig. 1).
The focal length of the camera is set as $f=0.16 \text{ m}$ and the scaling factor of the camera $\beta=1200.0$. Assume that the three axes of the camera frame which are denoted by $X_C$, $Y_C$ and $Z_C$, respectively are aligned with the axes $Y_0$, $Z_0$, and $X_0$ of the manipulator base frame, respectively, and the origins of the two frames has an offset along the axis $Z_C$, i.e., $d_C=6.0\text{ m}$. The lengths of the three links of the manipulator are set as $l_1=2.1 \text{ m}$, $l_2=2.1 \text{ m}$, and $l_3=1.9 \text{ m}$. 
The sampling period in the simulation is chosen to be 5 ms.

We first perform the simulation of the closed-loop system under the adaptive inverse-Jacobian-like control with the controller parameters being chosen as $K=40.0 I_3$, $\alpha=10.0$, $\Gamma_d=200.0I_8$, $\Gamma_z=0.008I_3$, $\Gamma_z^\perp=260.0I_2$. The initial values of the kinematic and camera parameter estimates are chosen as $\hat{l}_2(0)=\hat{l}_3(0)=3.2\text{ m}$, $\hat{d}_C(0)=3.2\text{ m}$, $\hat{f}(0)=0.09\text{ m}$, and $\hat{\beta}(0)=2000.0$. The initial value of the dynamic parameter estimate is chosen as $\hat{a}_d(0)=\left[0_6^T, 30, 0\right]^T$ while the actual value of the dynamic parameter is $a_d=[8.2688,
    2.9925,
    1.3538,$ $
    0.2578,
   10.6250,
    1.8050,
   46.3050,
   13.9650]^T$. The desired trajectory in the image space is given as
$
x_d=\begin{bmatrix}
53+21\cos(\pi t /3) \\
79+21\sin(\pi t /3)
\end{bmatrix}$.
The simulation results are plotted in Fig. 2, Fig. 3, and Fig. 4. As can be seen from Fig. 2, the image-space position tracking errors indeed converge to zero asymptotically. Fig. 3 illustrates the responses of the actual and estimated depths during the motion of the manipulator. It appears that the estimated depth has the tendency of tracking the actual depth, which is due to the depth parameter adaptation. Fig. 4 gives the response of the control torques.

We then perform the simulation of the closed-loop system under the adaptive transpose Jacobian control where the gain $K_1$ is chosen as $K_1=0.0015 I_2$, and the other controller parameters, the initial parameter estimates, and the desired image-space trajectory are chosen to be the same as above. The simulation results are shown in Fig. 5, Fig. 6, and Fig. 7.

The comparison between Fig. 2 and Fig. 5 and that between Fig. 3 and Fig. 6 show that the inverse-Jacobian-like control tends to yield better/smoother dynamic responses of the tracking errors and the estimated/actual depths than the transpose Jacobian control, yet from an overall perspective, their performance is comparable.


%
%
%

\begin{figure}
\centering
\begin{minipage}[t]{1.0\linewidth}
\centering
\includegraphics[width=2.8in]{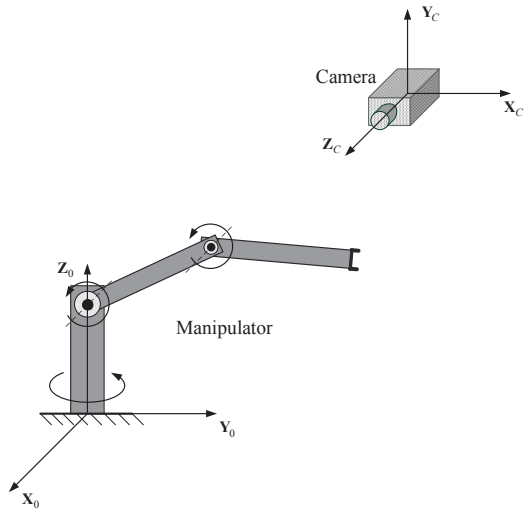}
\caption{Three-DOF manipulator with a fixed camera (from \cite{Wang2015_AUT})}\label{fig:side:a}
\end{minipage}%
\end{figure}


\begin{figure}
\centering
\begin{minipage}[t]{1.0\linewidth}
\centering
\includegraphics[width=2.5in]{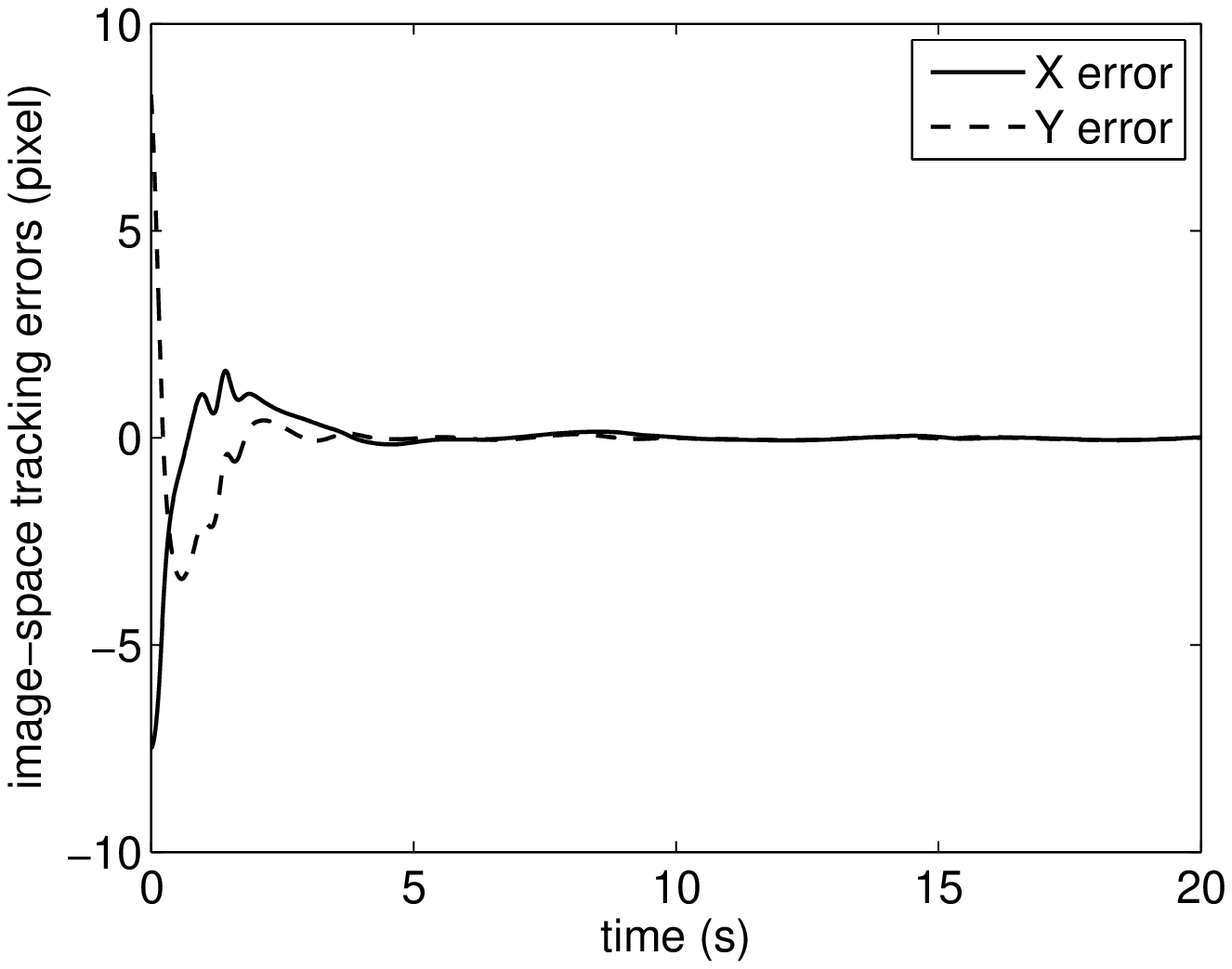}
\caption{Image-space position tracking errors}\label{fig:side:a}
\end{minipage}%
\end{figure}

\begin{figure}
\centering
\begin{minipage}[t]{1.0\linewidth}
\centering
\includegraphics[width=2.5in]{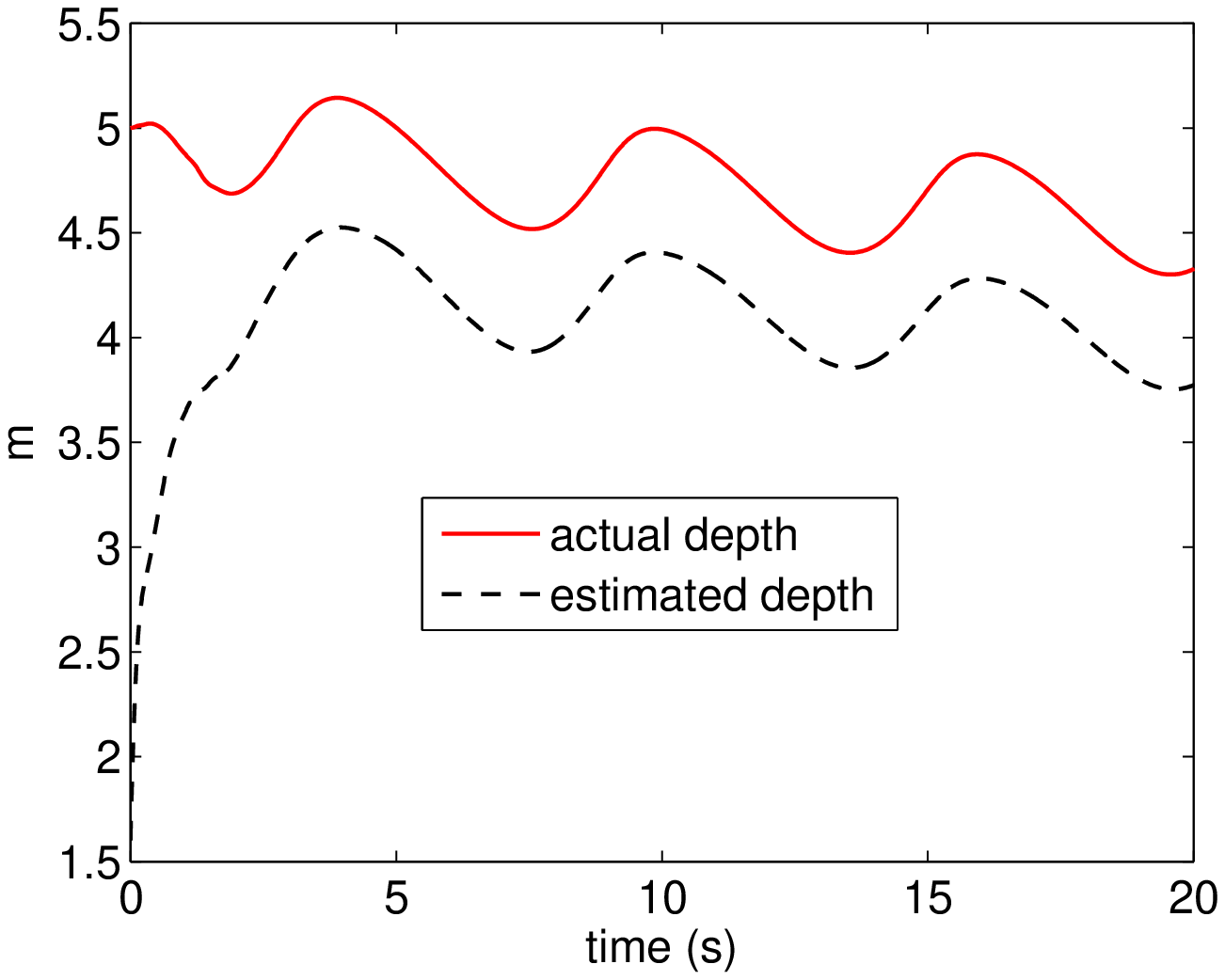}
\caption{Actual and estimated depths}\label{fig:side:a}
\end{minipage}%
\end{figure}

\begin{figure}
\centering
\begin{minipage}[t]{1.0\linewidth}
\centering
\includegraphics[width=2.5in]{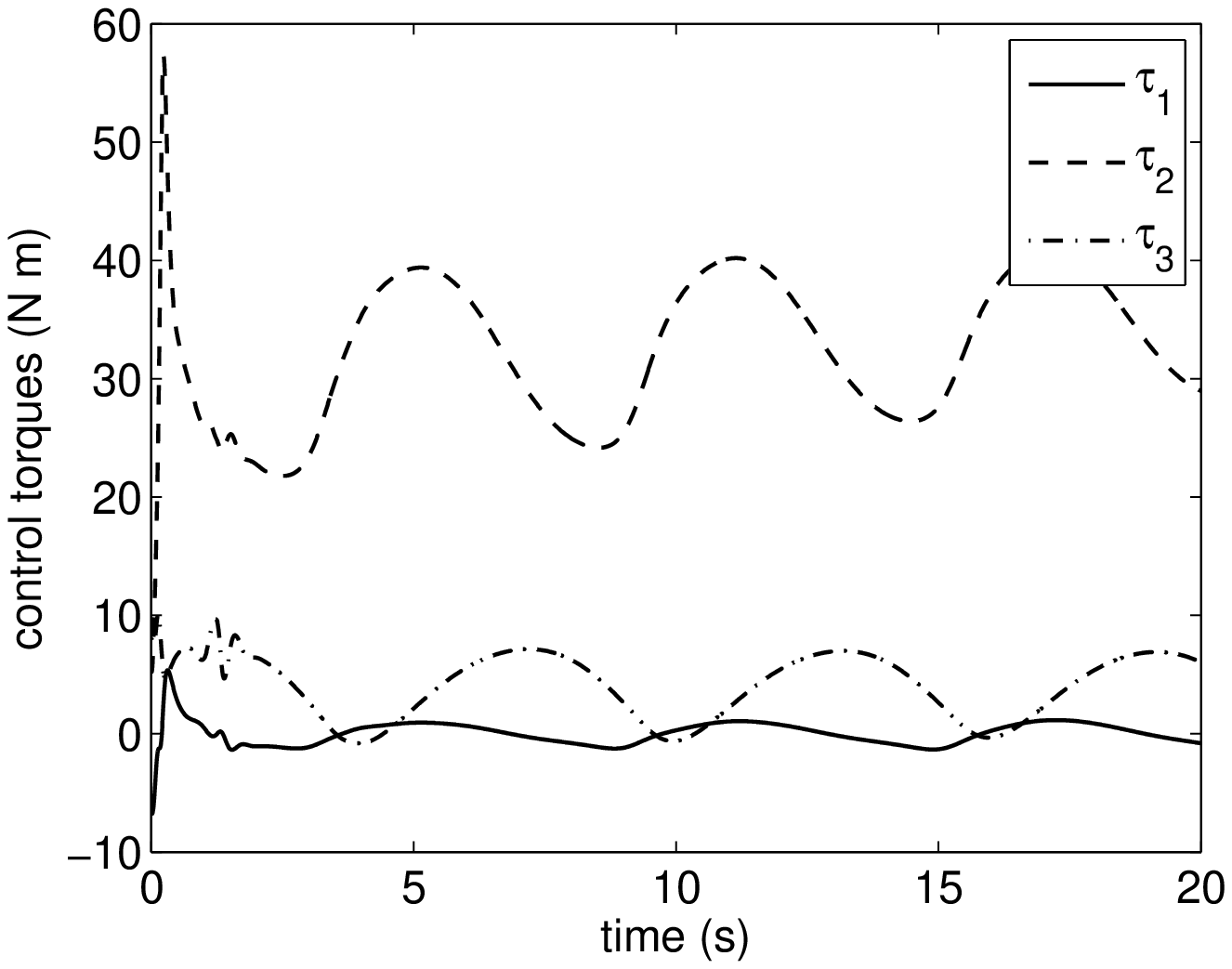}
\caption{Control torques}\label{fig:side:a}
\end{minipage}%
\end{figure}

\begin{figure}
\centering
\begin{minipage}[t]{1.0\linewidth}
\centering
\includegraphics[width=2.5in]{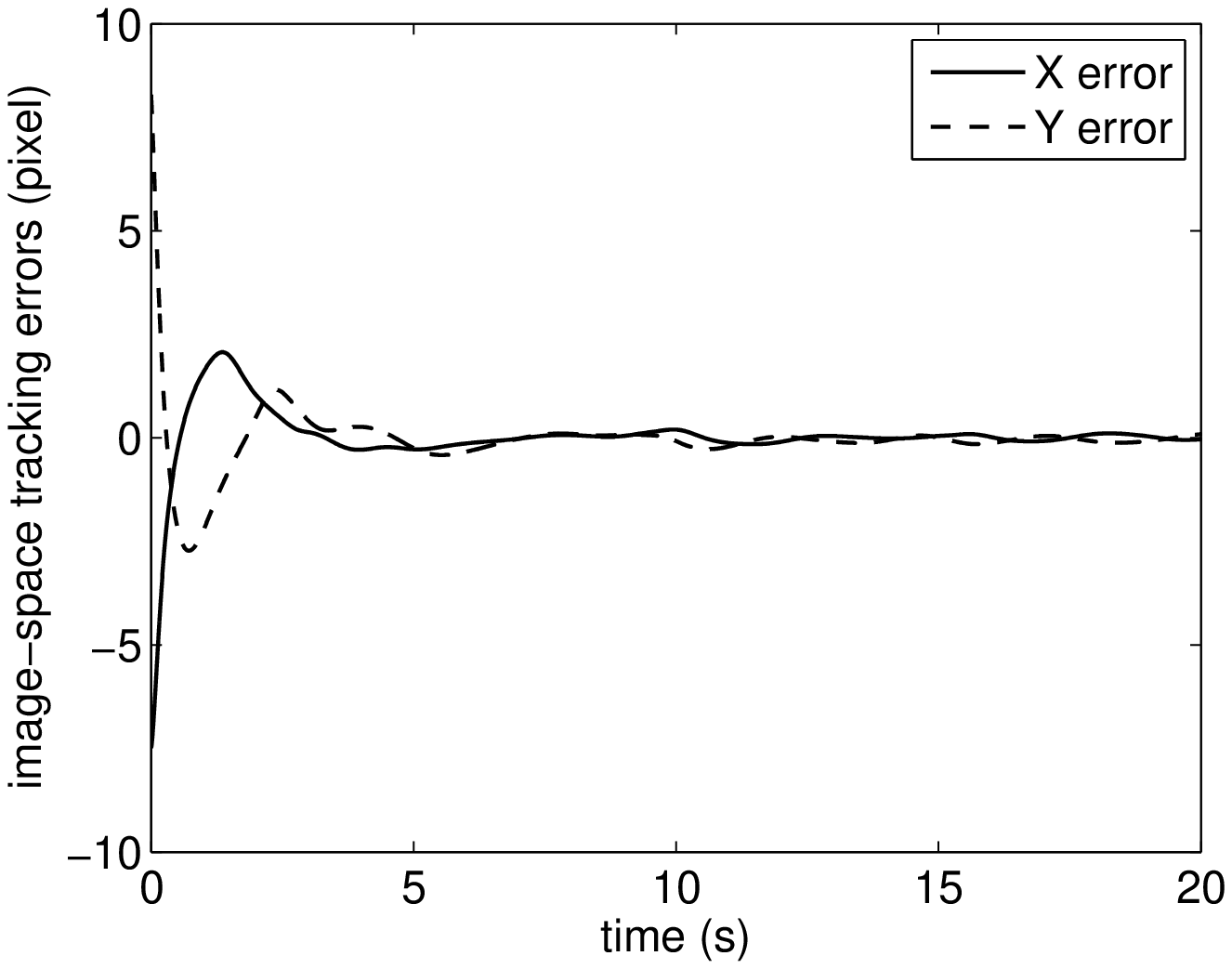}
\caption{Image-space position tracking errors}\label{fig:side:a}
\end{minipage}%
\end{figure}

\begin{figure}
\centering
\begin{minipage}[t]{1.0\linewidth}
\centering
\includegraphics[width=2.5in]{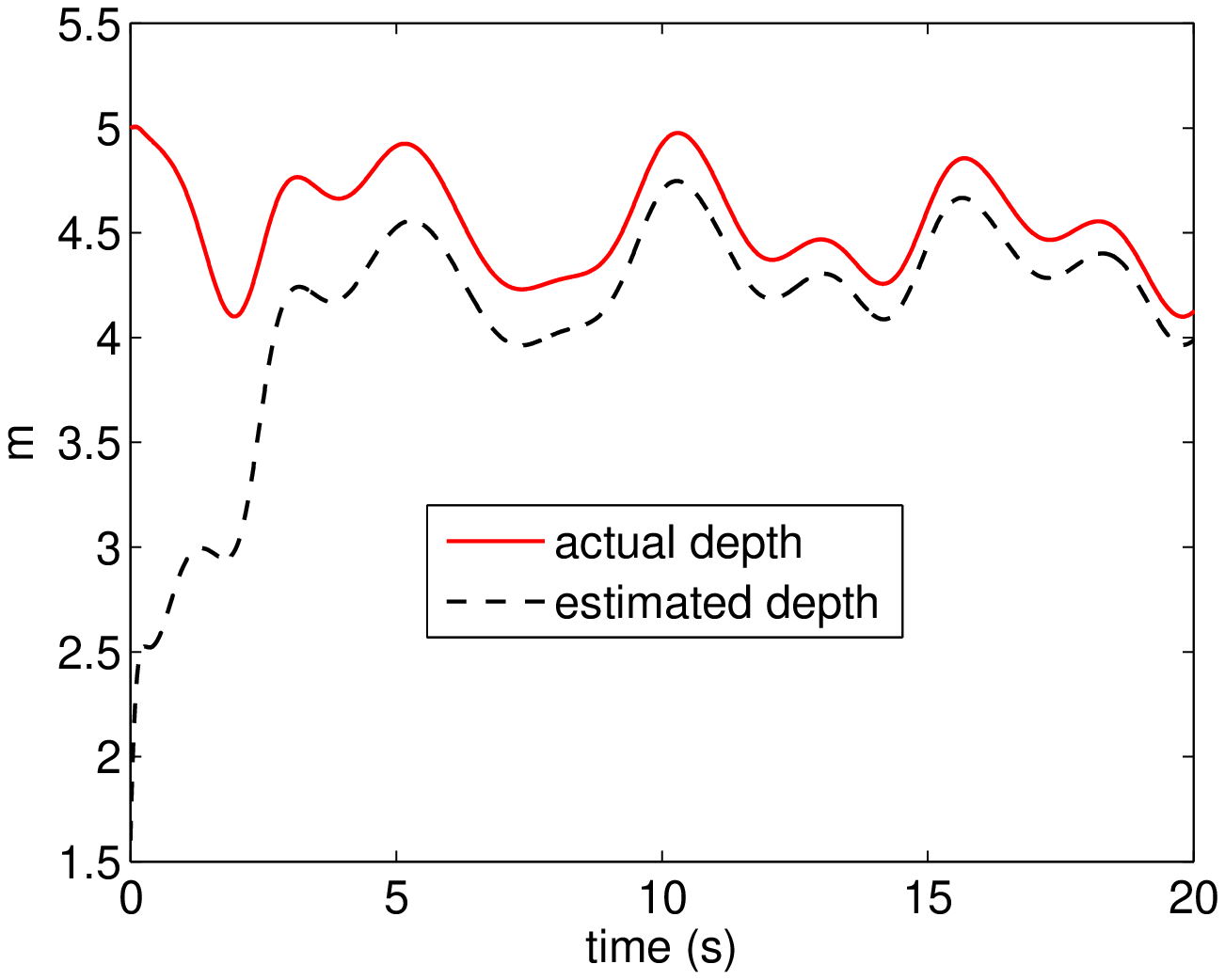}
\caption{Actual and estimated depths}\label{fig:side:a}
\end{minipage}%
\end{figure}

\begin{figure}
\centering
\begin{minipage}[t]{1.0\linewidth}
\centering
\includegraphics[width=2.5in]{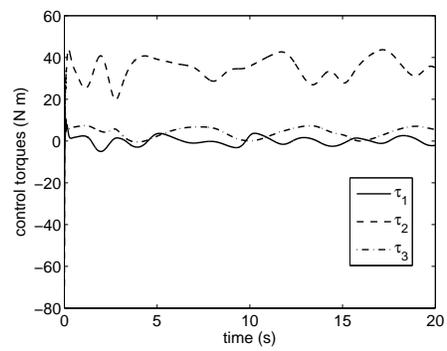}
\caption{Control torques}\label{fig:side:a}
\end{minipage}%
\end{figure}


%

\section{Conclusion}

In this paper, we have examined the tracking control problem for visually servoed robotic systems with uncertain kinematic, dynamic, and camera models. We start by formulating the passivity of the overall system kinematics, and then present two passivity-based adaptive control schemes. It is shown by the Lyapunov analysis approach that the image-space trajectory tracking errors converge to zero. It is also shown that one reduced version of the adaptive inverse-Jacobian-like controller is well suited to robots having an unmodifiable joint servoing controller yet admitting the design of the joint velocity command. Simulations using a three-DOF manipulator with a fixed camera are conducted to show the convergent property of the proposed adaptive controllers.

\bibliographystyle{IEEEtran}
\bibliography{..//Reference_list_Wang}

%
%
%

%








\end{document}